\documentclass[letterpaper, 10 pt, conference]{ieeeconf} 
\IEEEoverridecommandlockouts                              
\makeatletter

\let\proof\@undefined
\let\endproof\@undefined
\makeatother

\usepackage{cite}
\usepackage{amsmath} 
\usepackage{amssymb} 
\usepackage{amsthm}
\usepackage{dsfont}
\usepackage{mathrsfs}

\newtheorem{prop}{Proposition}

\newtheorem{fact}{Fact}
\newtheorem{rem}{Remark}

\newtheorem{thm}{Theorem}[section]
\newtheorem{lem}[thm]{Lemma}

\newcommand{\vect}[1]{\mathbf{#1}}
\newcommand{\mat}[1]{\mathbf{#1}}
\newcommand{\bs}[1]{\pmb{#1}}

\newcommand*\widebar[1]{\hbox{
    \vbox{%
      \hrule height 0.5pt 
      \kern0.5ex
      \hbox{%
        \kern-0.1em
        \ensuremath{#1}%
        \kern-0.1em
      }%
    }%
  }%
}

\newcommand{\colvec}[2][.9]{%
  \scalebox{#1}{%
    \renewcommand{\arraystretch}{.9}%
    $\begin{bmatrix}#2\end{bmatrix}$%
  }
}

\usepackage{color}
\usepackage{float}
\usepackage{subfigure}
\usepackage{graphicx}
\usepackage{caption}
\usepackage{hyperref}
\DeclareMathAlphabet\mathbfcal{OMS}{cmsy}{b}{n}

\usepackage[usenames,dvipsnames,table]{xcolor}

\usepackage{hyperref} 
\hypersetup{
    colorlinks,
    citecolor=black,
    filecolor=black,
    linkcolor=black,
    urlcolor=black,
    pdfauthor={},
    pdfsubject={},
    pdftitle={}
}
\usepackage{multicol}
\usepackage{multirow}

\title{\LARGE \bf {On the Distributed Estimation from Relative Measurements:\\ a Graph-Based Convergence Analysis}
}

\author{Marco Fabris, Giulia Michieletto and Angelo Cenedese
	\thanks{M. Fabris, G. Michieletto, A. Cenedese are with the Dept. of Information Engineering,
		University of Padova, Italy.
Corresponding author: M. Fabris {\tt\small marco.fabris.7@phd.unipd.it}
.}
	\thanks{The research leading to these results was partially supported by the MIUR Project SEAL - Smart\&safe Energy-aware Assisted Living (SCN-00398) and by the University of Padova grant BIRD168152.}
}

\begin{document}
	\maketitle
	\thispagestyle{empty}
	\pagestyle{empty}
	
	\begin{abstract}
For a multi-agent system state estimation resting upon noisy measurements constitutes a problem related to several application scenarios. 
Adopting the standard least-squares approach, in this work we derive both the (centralized) analytic solution to this issue and two distributed iterative schemes, which allow to establish a connection between the convergence behavior of consensus algorithm toward the optimal estimate and the theory of the stochastic matrices that describe the network system dynamics. 
This study on the one hand highlights the role of the topological links that define the neighborhood of agent nodes, while on the other allows to optimize the convergence rate by easy parameter tuning. 
The theoretical findings are validated considering different network topologies by means of numerical simulations.
\end{abstract}

\section{Introduction}
	

Many applications involving networked multi-agent systems lead to estimation problems that require the determination of the system state, interpreted as a set of scalar attributes associated to each element in the group. In this context, noisy measurements of the difference between certain pairs of them are generally assumed to be available. 
This is, for instance, the case of the orientation estimation for teams of ground robots equipped with bearing sensors~\cite{borra2013asynchronous}, 
the time-synchronization for sensor networks wherein the local clocks of each pairs of devices differ by a constant offset~\cite{barooah2007estimation},  
the statistical ranking in databases whose items can be assessed in a comparative manner~\cite{jiang2011statistical}, 
the branch-current-based state estimation in smart distribution system~\cite{pau2013efficient}, to cite a few.

All the aforementioned tasks can be framed as \emph{state estimation problems on a networked system given a set of noisy measurements}, typically relative measurements along a subset of network edges. 
A common and effective approach to face this issue turns out to be the least-squares (LS) optimization framework. 
Nonetheless, given the networked nature of multi-agent systems, the quest for distributed solutions becomes practically mandatory and of some research interest, and in addition, the convergence to a suitable optimum - as close as possible to the centralized LS solution - strictly depends on the topological features of the system. 


\noindent\textbf{Related works -} 
Several existing works explore the properties  of the distributed state estimation solution depending on the graph modelling the network in terms of available measurements. For example, in~\cite{rossi2017distributed}, the estimation task is studied 
proving that 
the system underlying topology determines whether the error of the LS optimal estimator decreases to zero as the number of unknown variables grows to infinity. Similarly, the authors of~\cite{barooah2009eerror} 
investigates how the variance of the estimation error of a node variable grows with its distance to an arbitrary reference one, providing then a graphs classification according to their sparsity. In~\cite{carron2014asynchronous}
 the problem of optimal estimating the position of each agent in a network from relative noisy distances is solved through a consensus-based algorithm whose rate of convergence is  determined for regular graphs such as Cayley, Ramanujan, and complete graphs. Finally, in~\cite{Tai2013} a distributed iterative estimation algorithm is proposed, where the iteration step is equal to the diameter
of the graph modeling a power network.

\noindent\textbf{Contributions -} 
The focus of this paper is the state estimation of a multi-agent system, by assuming that each agent in the group is able to retrieve relative noisy measurements w.r.t. some neighbors and also to communicate with them. In accordance with the current state of the art, the proposed solution exploits the LS optimization framework through the definition of a cost function based on the available measurements. 
The novel contribution consists in the analysis of the emergent connection between the minimization of such a convex functional and the theory of certain stochastic matrices, with the twofold aim of $i)$ providing an insight on the role of the topological links that define the neighborhood of an agent node and $ii)$ suggesting the design of distributed solutions whose performance can be easily tuned through a single network-related parameter. 
In this perspective, an iterative estimation procedure through two different update rules is proposed and the achieved results are compared with the centralized solution whose existence is guaranteed by the convexity of the cost functional. The convergence properties of the proposed 
update rules are theoretically discussed and 
then asserted by numerical results that account for different network topologies. 

\noindent\textbf{Paper structure -} Sec.~\ref{sec:graph_model} provides some notions on the graph-based multi-agent system model. In Sec.~\ref{sec:problem_solution} the optimal state estimation is formulated as a minimization problem and its centralized and distributed solutions are derived. Sec.~\ref{sec:convergence_analysis} is devoted to the convergence analysis of the proposed solutions. The theoretical observations are then validated by means of numerical simulations in Sec.~\ref{sec:simulation_results}. Finally, in Sec.~\ref{sec:conclusions} some conclusions are drawn.


\section{Graph-Based Network Model}
\label{sec:graph_model}
	
According to the existing literature, a $n$-agent system can be modeled through a graph $\mathcal{G}=\left(\mathcal{V},\:\mathcal{E}\right)$ so that each element in the nodes set $\mathcal{V}=\left\{v_1 \dots v_n\right\}$ corresponds to an agent in the group, while the edge set $\mathcal{E}\subseteq \mathcal{V}\times \mathcal{V}$ describes the agents interactions. 
In the rest of the paper, we assume that $\mathcal{G}$ is connected and undirected, and that set $\mathcal{E}$ depicts both agents sensing and communication capabilities, meaning that there exists $e_{ij}=(v_i , v_j) \in \mathcal{E}$ if and only if the $i$-th and $j$-th
agents can sense each other and are able to
reciprocally exchange information according to some predetermined communication protocol.

The agents interplays are generally represented by the \textit{adjacency matrix} $\mathbf{A} \in \mathbb{R}^{n \times n}$ such that $[\mathbf{A}]_{ij}=1$ if $e_{ij}\in\mathcal{E}$ ($v_i$ and $v_j$ are adjacent) and $[\mathbf{A}]_{ij}=0$ otherwise.
For each node $v_i$ in $\mathcal{G}$, the set $\mathcal{N}_i=\left\{v_j\in\mathcal{V}\;|\; [\mathbf{A}]_{ij}=1\right\} \subseteq \mathcal{V}$, named \textit{neighborhood}, thus represents the set of agents interacting with the $i$-th agent. By convention, it holds that $v_i \notin \mathcal{N}_i$. The 
cardinality of $\mathcal{N}_i$ is the \textit{degree}, $\text{deg}(v_i)$, of the $i$-th agent. This corresponds to the $i$-th element in the main diagonal of the \textit{degree matrix} $\mathbf{D} = \text{diag}(\mathbf{A}\mathds{1}_n)\in \mathbb{R}^{n\times n}$, where $\mathds{1}_n$ indicates a $n$-dimensional (column) vectors whose entries are all ones. The matrix
$\mathbf{D}$ in turn contributes to the definition of the \textit{Laplacian matrix} of $\mathcal{G}$, $\mathbf{L} = \mathbf{D}-  \mathbf{A} \in \mathbb{R}^{n \times n}$.

Finally, given a graph $\mathcal{G}$, a \textit{path} $\ell_{ij}$ from $v_i$ to $v_j$ consists of the adjacent nodes sequence $\ell_{ij} = \{v_i \ldots v_j \} \subseteq \mathcal{V}$, a cycle is a path whose starting and ending nodes coincide and the network \textit{diameter} is given by $\phi = \max\{\min (\vert \ell_{ij} \vert -1) \} \in \mathbb{R}$.

\section{State Estimation}
	\label{sec:problem_solution}
	
In this section, we first formalize the multi-agent system state estimation from relative measurements stating its centralized solution. Then, we derive two iterative schemes that solve the same issue by adopting a distributed paradigm.

\subsection{Problem Statement}
\label{sec:theOptimizationProblem}

Let consider a multi-agent system made up of $n$ devices whose interactions are represented by the undirected graph $\mathcal{G}$ according to the model described in Sec.~\ref{sec:graph_model}. We assume that each $i$-th agent in the network is associated to a scalar attribute $x_i \in \mathbb{R}$ (in the following also referred as \textit{$i$-th agent state}) corresponding to a certain physical quantity, and to a set of (noisy) relative measurements of the same quantity, namely to $\{\tilde{x}_{ij} \!\in \!\mathbb{R}, \; v_j \in \mathcal{N}_i\}$ where $\tilde{x}_{ij}$ represents a (noisy) measure of the difference between $x_j$ and $x_i$. 

The problem that we aim at solving concerns the determination of the  network configuration, i.e., the estimation of the set $\{x_1^\ast \ldots x_n^\ast \}$ that allows to approximate (and be consistent with) the set of the existing measurements. 
Formally, adopting the standard LS approach, we account for  the following optimization program 
\begin{align}
\label{eq:cost}
\underset{\{x_1 \ldots x_n\}}{\arg \min} & \; \frac{1}{2} \textstyle\sum\limits_{v_i \in \mathcal{V}} \textstyle\sum\limits_{v_j \in \mathcal{N}_i} ( x_i-x_j+\tilde{x}_{ij} )^2.
\end{align}

The cost function introduced in~\eqref{eq:cost}, hereafter denoted by $\varphi =\varphi(x_1 \ldots x_n)$, has a standard standard quadratic form, i.e., the minimization problem is convex. 
The existing literature about this issue is wide, and the novel  contribution of this work relies on the analysis of the existence of a distributed solution depending on the network topology, namely on the spectral properties of the graph $\mathcal{G}$ that also  influence the convergence behavior of $\varphi$ as discussed in~Sec~\ref{sec:convergence_analysis}. 

\subsection{Centralized vs Distributed Solution}
\label{sec:centralized_distributed_solution}

To derive a centralized solution for the minimization~\eqref{eq:cost}, we compute the gradient of $\varphi$ whose $i$-th component results
\begin{align}
[\bs{\nabla}_{\varphi}]_i = 
2\text{deg}(v_i)x_i - 2 \!\!\textstyle\sum\limits_{v_j \in {\mathcal N}_i}\!\! x_j - \!\!\textstyle\sum\limits_{v_j \in {\mathcal N}_i}\!\!(\tilde{x}_{ji} - \tilde{x}_{ij}).
\label{eq:costGraSO2}
\end{align}
As a consequence, the assignment $\bs{\nabla}_{\varphi} = \mathbf{0}_{n}$, where $\mathbf{0}_n \in \mathbb{R}^n$ indicates a $n$-dimensional (column) vectors whose entries are all zeros, yields the system
\begin{align}\label{eq:systemBetaDeltaBeta}
2 \mathbf {L}\vect{x} = \tilde{\vect{x}},
\end{align}
where $\tilde{\vect{x}}  = \colvec{ \textstyle\sum\nolimits_{v_j \in {\cal N}_1}(\tilde{x}_{j1} -\tilde{x}_{1j}) \;\; \ldots \;\; \textstyle\sum\nolimits_{v_j \in {\cal N}_n}(\tilde{x}_{jn} - \tilde{x}_{nj})}^\top \in\mathbb{R}^n$ and $\vect{x} = \colvec{ x_1 \;\; \ldots \;\;  x_n}^\top \in \mathbb{R}^n$. 	Hence, the next lemma can be straightforwardly obtained.
\begin{lem}[Centralized solution]
\label{Lemma1} 
Given a $n$-agents network associated to graph $\mathcal{G}$, if the relative
noisy measurements are such that $\tilde{\vect{x}} \notin \ker(\mathbf{L})\backslash\{\mathbf{0}_n\}$, the minimum norm solution of~\eqref{eq:cost} is given by
\begin{equation}
\vect{x}^\ast= \frac{1}{2} \mathbf{L}^{\dagger} {\tilde{\vect{x}}}, 
\label{eq:centrSol}
\end{equation} 
where $\mathbf{L}^\dagger\in\mathbb{R}^{n \times n}$ is the pseudo-inverse of the Laplacian matrix related to the graph $\mathcal{G}$.		
\end{lem}
%

Eq.~\eqref{eq:centrSol} represents the \textit{centralized} solution to the optimization~\eqref{eq:cost}: its computation requires to know both the network topology and the noisy relative measurements. On the other hand, each  equation of system~\eqref{eq:systemBetaDeltaBeta} refers to a single agent information, in terms of local topology and measurements, thus suggesting a \textit{distributed} approach. Formally, imposing $[\bs{\nabla}_\varphi ]_i = 0$ with $i \in \{1 \ldots n\}$, it follows that
\begin{equation}
x_i =  \frac{1}{\text{deg}(v_i)}\left(\textstyle\sum\limits_{v_j \in {\cal N}_i} x_j + \frac{1}{2} \!\! \textstyle\sum\limits_{v_j \in {\cal N}_i}(\tilde x_{ji} - \tilde x_{ij})\right),
\label{eq:distrUpdateScalar}
\end{equation}
i.e., the $i$-th node state depends exclusively on the states of its neighbors and on its set of relative measurements.

By considering the whole state vector $\vect{x}$, a discrete time system $\Sigma_{0}$ of the form
\begin{equation}
\Sigma_{0}: \quad \vect{x}({t+1}) = \mathbf{F}_{0} \vect{x}({t}) + \mathbf{u}_{0}
\label{eq:distrUpdateVector1}
\end{equation}
can be derived from~\eqref{eq:distrUpdateScalar} as an update rule driven by the input measurements. According to~\eqref{eq:distrUpdateVector1} the $i$-th state at the \mbox{$t$-th} step affects the neighbors estimates at the $(t+1)$-th step, but it is not considered for the recursive self-estimate. The state matrix $\mathbf{F}_{0} \in \mathbb{R}^{n \times n}$ is indeed equal to the adjacency matrix $\mathbf{A}$ normalized by the node degrees and it is thus a row-stochastic matrix. Formally, it occurs that
\begin{align}
\label{eq:matrix_F}
\mathbf{F}_{0}=\mathbf{D}^{-1} \mathbf{A}.
\end{align}
Similarly, the input $\mathbf{u}_{0} \in \mathbb{R}^n$ in~\eqref{eq:distrUpdateVector1} is given by the vector of the normalized relative measurements
\begin{equation}
\mathbf{u}_{0} = \frac{1}{2}\mathbf{D}^{-1}\tilde{\vect{x}}.
\label{eq:inputvector}
\end{equation}
From~\cite{landau1981bounds}, 
we can observe that if the graph $\mathcal{G}$ representing the given system is connected, then the state matrix $\mathbf{F}_{0}$ has $n$ real eigenvalues $\lambda^{\mathbf{F}_{0}}_0 \geq \ldots \geq \lambda^{\mathbf{F}_{0}}_{n-1}$ in the range $\left[-1,1\right]$ with $\lambda^{\mathbf{F}_{0}}_0=1$ having single algebraic multiplicity.
	%
	
	%
	
\subsection{Distributed Spectral Based Solution} \label{subsec:alg1}

A different distributed model can be constructed by introducing some memory in the system and adopting~\eqref{eq:distrUpdateVector1} to provide only a weighted correction to current estimate, thus leading to
\begin{align}
\Sigma_{\eta}: \quad \vect{x}(k+1) 	&= \eta\vect{x}(k)+(1-\eta)\left(\mathbf{F}_{0}\vect{x}(k) + \mathbf{u}_{0}\right) \nonumber\\
&= \left(\eta \mathbf{I}_n +(1-\eta)\mathbf{F}_{0}\right)\vect{x}(k)+(1-\eta)\mathbf{u}_{0} \nonumber\\ 
& = \mathbf{F}_{\eta} \vect{x}(k) + \vect{u}_\eta, \label{eq:distrUpdateVector_eta}
\end{align} 
where $\eta \in (0,1)$ and $\mathbf{I}_n$ is the identity matrix of dimension $n$. The state matrix $\mathbf{F}_{\eta} \in \mathbb{R}^{n \times n}$ is still row-stochastic but with eigenvalues in the range $\left(-1+2\eta,1\right]$. In particular, we can observe that, exploiting the linearity of the spectrum, it holds that $\lambda_{i}^{\mathbf{F}_{{\eta}}} = \eta + (1-\eta)\lambda_{i}^{\mathbf{F}_{0}}$, $i=0 \ldots n-1$.

It can be shown that if the equilibrium points of systems \eqref{eq:distrUpdateVector1} and~\eqref{eq:distrUpdateVector_eta} exist, these are those stated by Lemma~\ref{Lemma1}. In other words, these equilibria represent the \textit{distributed} solutions of problem~\eqref{eq:cost}.

\section{Convergence Analysis}
	\label{sec:convergence_analysis}

To study the state estimation behavior, it is necessary to distinguish between the convergence of the cost function $\varphi$ toward zero and the convergence of the states to the equilibrium values, being the former a function of the others. 
It is desirable that the update schemes~\eqref{eq:distrUpdateVector1} and~\eqref{eq:distrUpdateVector_eta} guarantee a monotonic decrease of the cost function, e.g., by adopting a gradient descent  procedure. However, it appears that the convergence of the states toward the equilibrium configuration depends on the eigenvalues of the state matrices of the evaluated model ($\mathbf{F}_{0}$ and $\mathbf{F}_{\eta}$) and, in particular, the convergence rate depends on the second largest eigenvalue~in~modulus.
	
\subsection{State-Space Model $\Sigma_{0}$} 

Considering system~\eqref{eq:distrUpdateVector1}, whose dynamics is governed by $\mathbf{F}_{0}$, the states convergence is ensured only if $\lambda=-1$ is not among the eigenvalues of the  matrix. 
Indeed, if $\lambda=-1$ belongs to the spectrum of $\mathbf{F}_{0}$ its multiplicity would be unitary 
and this would imply (for large observation times $t\!\gg\!1$) constant oscillations of the states around their equilibrium values due to the presence of a dominant oscillatory mode. From a graphical model point of view, this case can be interpreted referring to the network topology since $\lambda=-1$ occurs if and only if the associated graph is \textit{bipartite}, i.e., all its cycles consist of an even number of nodes~\cite{landau1981bounds}. This statement can be proved accounting for the next fact~\cite{chung1997spectral}.

\begin{fact}
Denoting by $\bs{\varpi}_i \in \mathbb{R}^n$ the $i$-th eigenvector of $\mathbf{F}_{0}$, 
 we consider the following change of variables  $\bs{\varpi}_i^\prime = \mat{D}^{1/2}\bs{\varpi}_i$.
Because of the spectrum linearity, for $i \in \{0\ldots n-1\}$, it holds that
\begin{align} 
 \mathbf{F}_{0} \, \mat{D}^{-1/2}\bs{\varpi}_i^\prime = \lambda^{\mathbf{F}_{0}}_{i} \mat{D}^{-1/2}\bs{\varpi}_i^\prime \; \Rightarrow \;
 {\mathbfcal{L}}\;\bs{\varpi}_i^\prime = (1-\lambda^{\mathbf{F}_{0}}_{i}) \bs{\varpi}_i^\prime,
	\end{align}
where $\mathbfcal{L}=\mathbf{I}_n-\mat{D}^{-1/2} \mathbf{A} \mat{D}^{-1/2} \in \mathbb{R}^{n \times n}$ is the \textit{normalized Laplacian matrix} of $\mathcal{G}$.
Hence, indicating with $\lambda_{i}^{\mathbfcal{L}}$ the eigenvalues of $\mathbfcal{L}$, such that $0=\lambda_{0}^{\mathbfcal{L}} < \lambda_{1}^{\mathbfcal{L}} \leq \dots \leq \lambda_{n-1}^{\mathbfcal{L}} \leq 2$, we have that
\begin{equation}
\label{eq:correspondance_between_lapl_and_F_eigs_}
\lambda_{i}^{\mathbfcal{L}}=1-\lambda_{i}^{\mathbf{F}_{0}}, \quad \text{for } i=0 \ldots n-1.
	\end{equation}
Moreover, $\lambda_{n-1}^{\mathbfcal{L}} = 2$ if and only  if $\mathcal{G}$ is bipartite, and consequently, $\lambda_{n-1}^{\mathbf{F}_{0}} = -1$ having unitary multiplicity. 
\end{fact}	
	
When the graph describing the network is bipartite, convergence to the equilibrium can be reached only if it is possible to modify such topology so that the resulting graph presents at least one cycle made of an odd number of nodes. For example, in the multi-agent scenario, the link addition case (corresponding to a graph topology modification) translates into the possibility of finding an additional connection among devices that are neighbors in the sense of both communication and sensing: such an edge selection solution may not be feasible or convenient in real-world~applications.

Conversely, in the non-bipartite case, convergence is assured and its rate is governed by the second largest (in modulus) eigenvalue of the state matrix $\mathbf{F}_{0}$.

\subsection{State-Space Model $\Sigma_{\eta}$} 
\label{sec:convergence_analysis_eta}
	
The presence of self-loops in~\eqref{eq:distrUpdateVector_eta}, controlled by parameter $\eta$, allows to modify the eigenvalues domain from the unit circle to the set $\Upsilon \cup \{1\}$ where $\Upsilon$ is a circle centered in $(\eta,0)$ with radius $1-\eta<1$, ruling out the possible presence of the critical eigenvalue $\lambda = -1$.
	Hence, not only the stability of the system is obtained but also the states convergence is always assured.
	More interestingly, the $\eta$ parameter can be tuned to control the convergence speed, governed by the second largest (in modulus) eigenvalue of~$\mathbf{F}_{\eta}$. If this is negative the estimated states converge toward the equilibrium values through an oscillatory transient period. 
	On the contrary, if the second largest eigenvalue is positive then the estimation trend is asymptotically convergent, monotonic for large observation times. 
	From an applicative perspective, the former oscillatory behavior may be preferable since an averaging operation might provide an approximate solution to the convergence value, while in the latter case the iterations might consistently underestimate or overestimate the asymptotic values.
	Remarkably, these different behaviors can be seen as dependent on the control parameter $\eta$.
	
A good and viable strategy is to select the parameter $\eta$ as
\begin{equation}\label{eq:min_prob_eta_star}
\eta^\ast = \underset{\eta \in [0,1)}{\arg \min} \left\lbrace  \underset{i=1\ldots n-1}{\max}\left|\lambda_{i}^{\mathbf{F}_{\eta}}\right|  \right\rbrace,
\end{equation}
to minimize the convergence rate of scheme $\Sigma_{\eta}$. The value of  $\eta^\ast$ can be analytically computed as shown in the next proposition. Note that in~\eqref{eq:min_prob_eta_star} we assume that $\eta$ ($\eta^\ast$) might be zero: this case corresponds to consider model~\eqref{eq:distrUpdateVector1}.

\begin{prop}\label{prop:eta_opt}
Given a multi-agent network represented by graph $\mathcal{G}$, the optimal value $\eta^\ast$ in~\eqref{eq:min_prob_eta_star} is univocally determined as
\begin{align}\label{eq:etaast}
\eta^\ast=\begin{cases}
1-\varsigma_\mathbfcal{L}^{-1}, \quad &\varsigma_\mathbfcal{L} > 1 \\
0, \quad &\varsigma_\mathbfcal{L} \leq 1 
\end{cases}
\end{align}
where $\varsigma_\mathbfcal{L} = \frac{1}{2}(\lambda_{1}^\mathbfcal{L} + \lambda_{n-1}^{\mathbfcal{L}}) \in \mathbb{R}$ and $\mathbfcal{L}$ is the normalized Laplacian matrix related to $\mathcal{G}$.
\end{prop}
\begin{proof}
Because \mbox{$\lambda^{\mathbf{F}_{\eta}}_1 \geq \ldots \geq  \lambda^{\mathbf{F}_{\eta}}_{n-1}$} for all $\eta \in [0,1)$ and the spectrum linearity, problem \eqref{eq:min_prob_eta_star} can be rewritten as
	\begin{align}
	\eta^\ast 
	&= \underset{\eta \in [0,1)}{\arg \min} \left\lbrace \max\left\lbrace  \left|\lambda_{1}^{\mathbf{F}_{\eta}}\right|, \left|\lambda_{n-1}^{\mathbf{F}_{\eta}}\right|  \right\rbrace \right\rbrace, \label{eq:argmineqtstar_minprobspecprop}
	\end{align}
where $\lambda_{1}^{\mathbf{F}_{\eta}}= \eta + (1-\eta)\lambda_{1}^{\mathbf{F}_{0}} $ and $\lambda_{n-1}^{\mathbf{F}_{\eta}}= \eta + (1-\eta)\lambda_{n-1}^{\mathbf{F}_{0}} $.
Let then distinguish two cases: $(a)$ $\varsigma_\mathbfcal{L} > 1$ and $(b)$ $\varsigma_\mathbfcal{L} \leq 1$.\\
$(a)$ The solution of~\eqref{eq:argmineqtstar_minprobspecprop} follows from the relation
\begin{align}\label{eq:linalgeqetastar}
& -\left(\eta + (1-\eta)\lambda_{n-1}^{\mathbf{F}_{0}}\right)  =  \eta + (1-\eta)\lambda_{1}^{\mathbf{F}_{0}}, 
\end{align}
under the constraint $\eta\in [0,1)$. Exploiting~\eqref{eq:correspondance_between_lapl_and_F_eigs_}, it results that
\begin{align}
\eta^\ast \!=\! \frac{\lambda_{1}^{\mathbf{F}_{0}} + \lambda_{n-1}^{\mathbf{F}_{0}}}{\lambda_{1}^{\mathbf{F}_{0}} + \lambda_{n-1}^{\mathbf{F}_{0}}-2} \!=\! \frac{2-(\lambda_{1}^{\mathbfcal{L}} + \lambda_{n-1}^{\mathbfcal{L}})}{-(\lambda_{1}^{\mathbfcal{L}} + \lambda_{n-1}^{\mathbfcal{L}})} \!=\! 1-\varsigma_\mathbfcal{L}^{-1} \label{eq:final_result_eta_star_case1}.
	\end{align}
$(b)$ If $\varsigma_\mathbfcal{L} \leq 1$ solution~\eqref{eq:final_result_eta_star_case1} is not valid because of the constraint on $\eta$.  Furthermore, problem~\eqref{eq:argmineqtstar_minprobspecprop} boils down to 
	\begin{align}\label{eq:minimzation_cas2etastar}
	\eta^\ast = \underset{\eta \in [0,1)}{\arg \min} \left\lbrace  \eta + (1-\eta)\lambda_{1}^{\mathbf{F}_{0}}  \right\rbrace,
	\end{align}
	because $\lambda_{1}^{\mathbf{F}_{0}}=1-\lambda_{1}^{\mathbfcal{L}}$,  $\lambda_{1}^{\mathbf{F}_{0}} \in [0,1)$, results
	to be the largest eigenvalue (in modulus) of $\mathbf{F}_{0}$  inside the unit circle.
	Since term $\eta + (1-\eta)\lambda_{1}^{\mathbf{F}_{0}}$ in \eqref{eq:minimzation_cas2etastar} is strictly increasing in the variable $\eta$, the minimization is attained for $\eta^\ast = 0$.
	\end{proof}

\begin{rem}	
Prop.~\ref{prop:eta_opt} shows that, when $\varsigma_\mathbfcal{L} \leq 1$, the solution of~\eqref{eq:min_prob_eta_star} is $\eta^\ast= 0$. Thus, in this case, either $\eta$ should be selected in the neighborhood of zero or the scheme $\Sigma_{0}$ should be considered optimal. 
\end{rem}

\begin{rem}
Note that, given a multi-agent network represented by a bipartite graph $\mathcal{G}$, the optimal value $\eta^\ast$ in~\eqref{eq:min_prob_eta_star} results to be
$
	\eta^\ast =\lambda_{1}^{\mathbfcal{L}}(2+\lambda_{1}^{\mathbfcal{L}})^{-1} \in (0,1).
$
\vspace{0.1cm}
\label{Rem: 3}
\end{rem}

The results in Prop.~\ref{prop:eta_opt} allows to provide a lower bound for the convergence rate $\mathfrak{r}_\eta =
{\max}_{i=1\ldots n-1}|\lambda_{i}^{\mathbf{F}_{\eta}}|$ of model $\Sigma_{\eta}$.  
If $\varsigma_\mathbfcal{L} \leq 1 $ then it trivially holds $\mathfrak{r}_\eta \ge \mathfrak{r}_{\eta^\ast} = 1-\lambda_{1}^{\mathbfcal{L}}$. 
Nevertheless, when $\varsigma_\mathbfcal{L} > 1$, it is possible to prove that the convergence rate depends on the diameter of the considered network. 
To do so, we first observe that, if $\varsigma_\mathbfcal{L} > 1$, by combining~\eqref{eq:linalgeqetastar} and~\eqref{eq:final_result_eta_star_case1}, it occurs 
	\begin{equation}\label{eq:etastar_cr_alone}
	\mathfrak{r}_\eta \ge \mathfrak{r}_{\eta^\ast} = 
	1-\frac{\lambda_{1}^{\mathbfcal{L}}}{\varsigma_\mathbfcal{L}} = 
	\frac{\lambda_{n-1}^{\mathbfcal{L}} -\lambda_{1}^{\mathbfcal{L}}}{\lambda_{n-1}^{\mathbfcal{L}}+\lambda_{1}^{\mathbfcal{L}}}.
	\end{equation}
Then, we recall the following inequality
valid for any non-complete graph $\mathcal{G}$ with $n$ vertices~\cite{chung1997spectral}
	\begin{equation}\label{ineq:chung_diameter}
	\phi \leq \left\lceil \frac{\mathrm{arcosh} (n-1)}{\mathrm{arcosh}\left( (\lambda_{n-1}^{\mathbfcal{L}}+\lambda_{1}^{\mathbfcal{L}})/ (\lambda_{n-1}^{\mathbfcal{L}}-\lambda_{1}^{\mathbfcal{L}})\right)} \right\rceil.
	\end{equation}
Combining~\eqref{eq:etastar_cr_alone} and~\eqref{ineq:chung_diameter}, a lower bound for the convergence rate can be derived depending only on topological properties:
\begin{equation}\label{ineq:lower_bound_conv_rate}
\mathfrak{r}_\eta \ge \mathfrak{r}_{\eta^\ast} \geq \mathrm{sech} \left(\frac{\mathrm{arcosh}(n-1)}{\phi-1}\right).
\end{equation}

Note that for a complete graph with $n$ vertices it holds that $\lambda_{1}^{\mathbfcal{L}} =\lambda_{n-1}^{\mathbfcal{L}} = n({n-1})^{-1}$, hence $\eta^\ast = {n}^{-1}$ and $\mathfrak{r}_{\eta^\ast} = 0$. This implies that the estimate procedure converges toward the equilibrium in one single step.

\begin{table*}
\centering
\renewcommand\arraystretch{1.2}
\begin{tabular}{c|c|c||c|c|}
\cline{2-5}
\multicolumn{1}{r|}{}  & \multicolumn{2}{c||}{$n$ odd} & \multicolumn{2}{c|}{$n$ even} \\ \cline{2-5}
\multicolumn{1}{r|}{}	& \multirow{1}{*}{exact value}      & $ n\rightarrow +\infty$    & \multirow{1}{*}{exact value}           & $n\rightarrow +\infty$        \\ \hline \hline
\multicolumn{1}{|c|}{$\lambda_{1}^{\mathbfcal{L}}$} & $8s^{2}_{\theta_n} c^{2}_{\theta_n} $  & $\dfrac{2 \pi^{2}}{n^{2}}$  & $8s^{2}_{\theta_n} c^{2}_{\theta_n} $ & $\dfrac{2 \pi^{2}}{n^{2}}$ \\ \hline
		\multicolumn{1}{|c|}{$\lambda_{n-1}^{\mathbfcal{L}}$} & $2c^{2}_{\theta_n} $ & $2-\dfrac{\pi^{2}}{2n^{2}}$  & $2$ & $2$  \\ \hline
		\multicolumn{1}{|c|}{$\sigma_\mathbfcal{L}$}& $c^{2}_{\theta_n} +4s^{2}_{\theta_n} c^{2}_{\theta_n} >1$ & $1+\dfrac{3\pi^{2}}{4n^{2}}$ &  $1+4s^{2}_{\theta_n} c^{2}_{\theta_n} >1$  & $1+\dfrac{\pi^{2}}{n^{2}}$  \\ \hline
		\multicolumn{1}{|c|}{$\eta^\ast$} & $\dfrac{4s^{2}_{\theta_n} c^{2}_{\theta_n} -s^{2}_{\theta_n} }{4s^{2}_{\theta_n} c^{2}_{\theta_n} +c^{2}_{\theta_n} }$ & $\dfrac{3}{4}\dfrac{\pi^{2}}{\pi^{2}+n^{2}}$  & $\dfrac{4s^{2}_{\theta_n} c^{2}_{\theta_n} }{1+4 s^{2}_{\theta_n}  c^{2}_{\theta_n} }$ & $\dfrac{\pi^{2}}{\pi^{2}+n^{2}}$   \\ \hline
		\multicolumn{1}{|c|}{$r_{\eta^\ast}$} & $\dfrac{1-4s^{2}_{\theta_n} }{1+4s^{2}_{\theta_n} }$ & $\dfrac{n^{2}-\pi^{2}}{n^{2}+\pi^{2}}$  & $\dfrac{1-4s^{2}_{\theta_n} c^{2}_{\theta_n} }{1+4s^{2}_{\theta_n} c^{2}_{\theta_n} }$ & $\dfrac{n^{2}-\pi^{2}}{n^{2}+\pi^{2}}$ \\ \hline
	\end{tabular}
	\caption{Evaluation of the quantities introduced in Sec.~\ref{sec:convergence_analysis_eta} for a ring topology network, using the following notation $s_{\theta_n}  = \sin(\theta_n)$, $c_{\theta_n} =\cos(\theta_n) $  and $\theta_n = \pi/(2n) \in (0,\pi/6]$.}
	\label{tab:ring}
\end{table*}

\section{Numerical Results}
	\label{sec:simulation_results}
	
To compare the two state-space update models~\eqref{eq:distrUpdateVector1} and~\eqref{eq:distrUpdateVector_eta} and validate the convergence properties discussed in the previous section, in the following we consider different case studies, involving some of the most popular topology types. 

\subsection{Analytical Results for a Ring Topology}

Firstly, we consider a system of $n\geq 3$ agents arranged according to a ring topology so that each device interacts with the two most closed ones. Labelling the agents in a suitable way, such an architecture is described by the following circulant adjacency matrix
\begin{equation}\label{eq:adjring}
[\mathbf{A}]_{ij}= \begin{cases}
1, \quad &j= n-\mathrm{mod}(n-i\pm 1,n)\\
0, \quad &\text{otherwise}
\end{cases}
\end{equation}
and the eigenvalues of the matrix $\mat{F}_0$ have the following analytic expression depending on $\theta_n=\pi/(2n)$~\cite{broughton2014analytical}
\begin{equation}
\label{eq:spectrumF}
\lambda_i^{\mat{F}_0} =  \cos\left(4\theta_n i \right), \ i=0 \ldots n-1. 
\end{equation}

From~\eqref{eq:spectrumF}, it is straightforward to note that  
$\lambda=-1$ belongs to the spectrum of $\mat{F}_0$ if and only if $n$ is even, i.e., if the graph is bipartite.  
Moreover, we observe that  the largest negative eigenvalue in modulus of $\mat{F}_0$ is $\lambda^{\mat{F}_0}_{{n}/{2}}$ when $n$ is even and $\lambda^{\mat{F}_0}_{{(n-1)}/{2}}$  when $n$ is odd, whereas the largest positive eigenvalue (excluding $\lambda^{\mat{F}_0}_0=1$) is always $\lambda^{\mat{F}_0}_1$. Hence, because of~\eqref{eq:correspondance_between_lapl_and_F_eigs_}, we have that
\begin{align}
\lambda_1^{\mathbfcal{L}} &= 2 \sin^2 \left( 2\theta_n \right) =8\sin^2\theta_n\cos^2\theta_n \label{eq:lambda_1} \\
\lambda_{n-1}^{\mathbfcal{L}} &= 
\begin{cases} 
2 &n \; \text{even}\\
1+\cos\left( 2\theta_n \right)=\ &n \; \text{odd}\\
\end{cases} \label{eq:lambda_n}
\end{align}

\begin{figure}[b!]
    \centering
{\includegraphics[width=1\columnwidth]{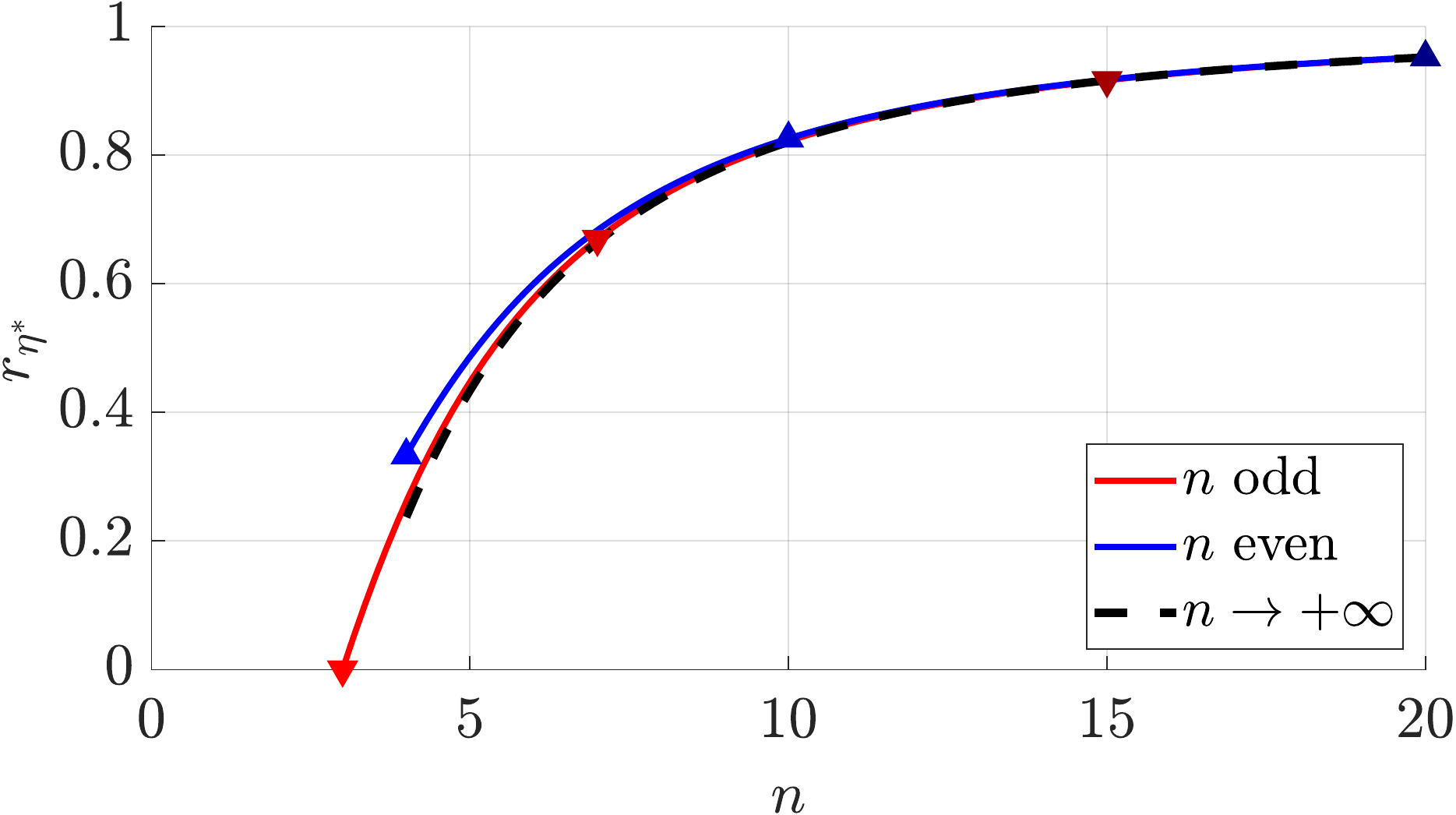}}
    \caption{Trend of $r_{\eta^\ast}$ w.r.t. number $n$ of devices composing the ring network \eqref{eq:adjring}. }
    \label{fig:sim_param_ring_a}
\end{figure}

Exploiting~\eqref{eq:lambda_1}-\eqref{eq:lambda_n}, it is possible to compute 
the values of the quantities introduced in Sec.~\ref{sec:convergence_analysis_eta} for such ring-graphs. 
These are reported in Tab.~\ref{tab:ring}. 
Observe that, as the number $n$ of devices increases, the convergence performance $r_{\eta^\ast}$ quadratically deteriorates (Fig.~\ref{fig:sim_param_ring_a}). 

	\begin{figure*}[t!]
		\centering
		\subfigure[Regular non-circulant topology]{\includegraphics[width=0.325\textwidth]{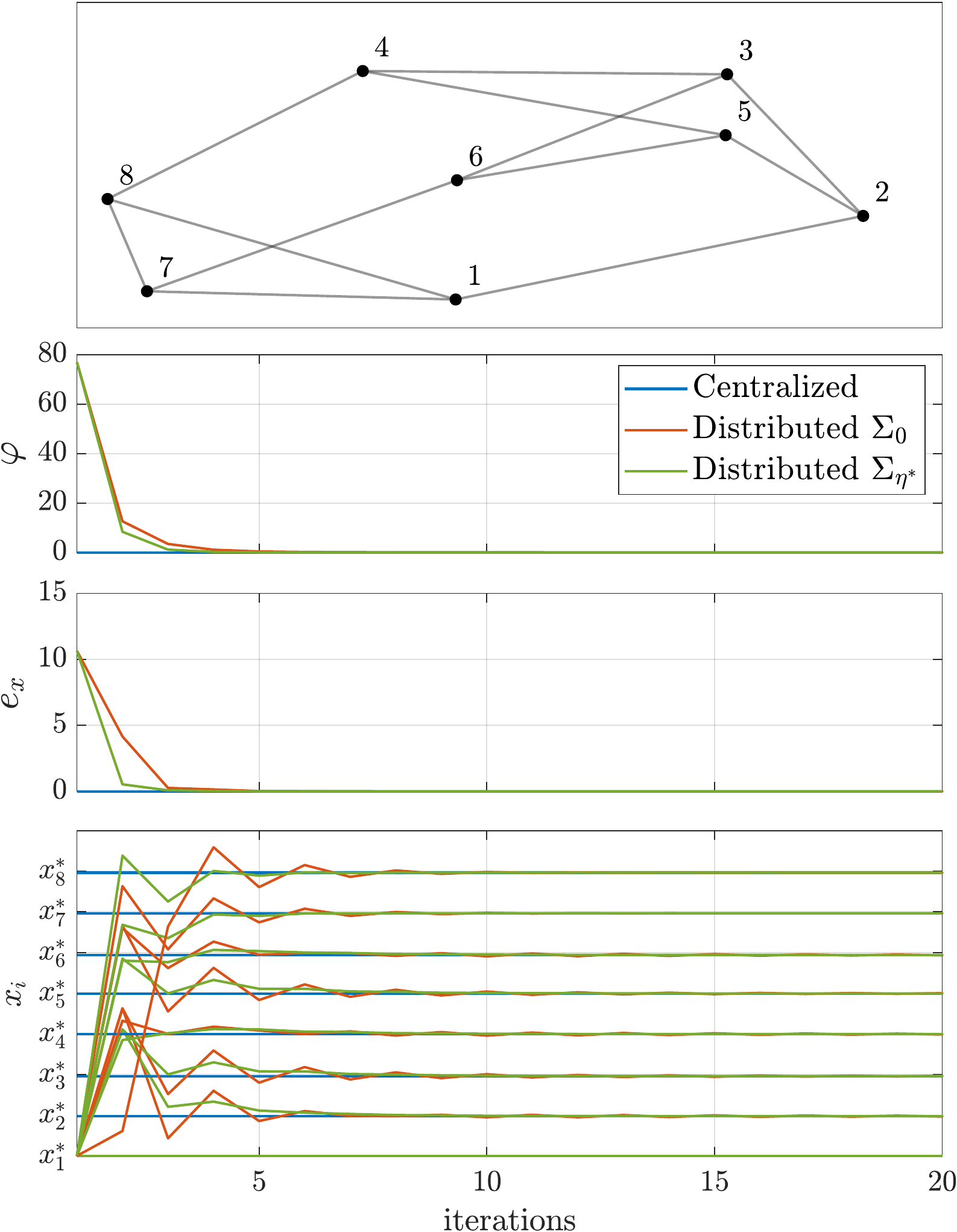}\label{fig:sim_1}}
		\hfill
		\subfigure[Non-regular bipartite topology]{\includegraphics[width=0.325\textwidth]{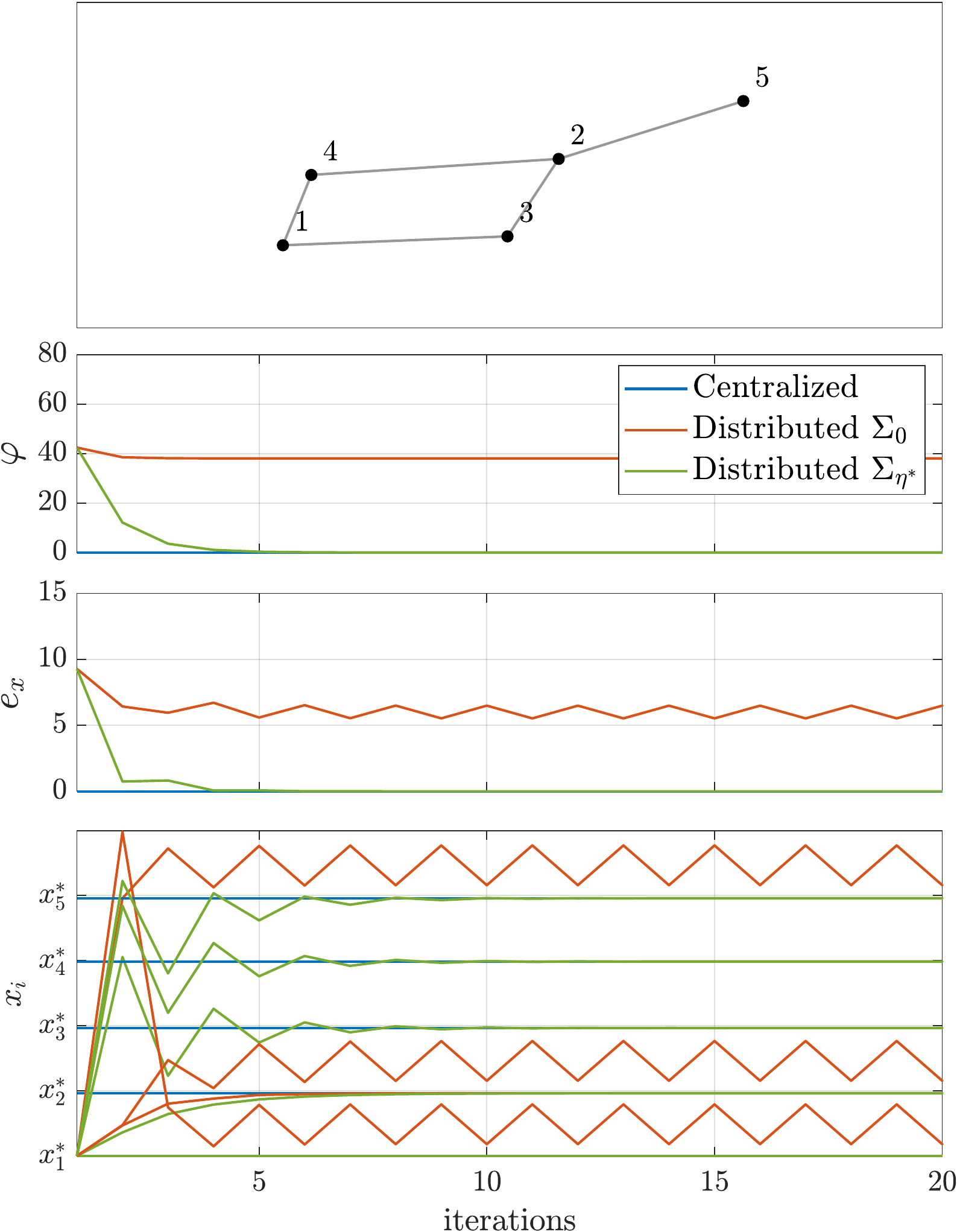}\label{fig:sim_2}}
		\hfill
		\subfigure[Non-regular small-world topology]{\includegraphics[width=0.325\textwidth]{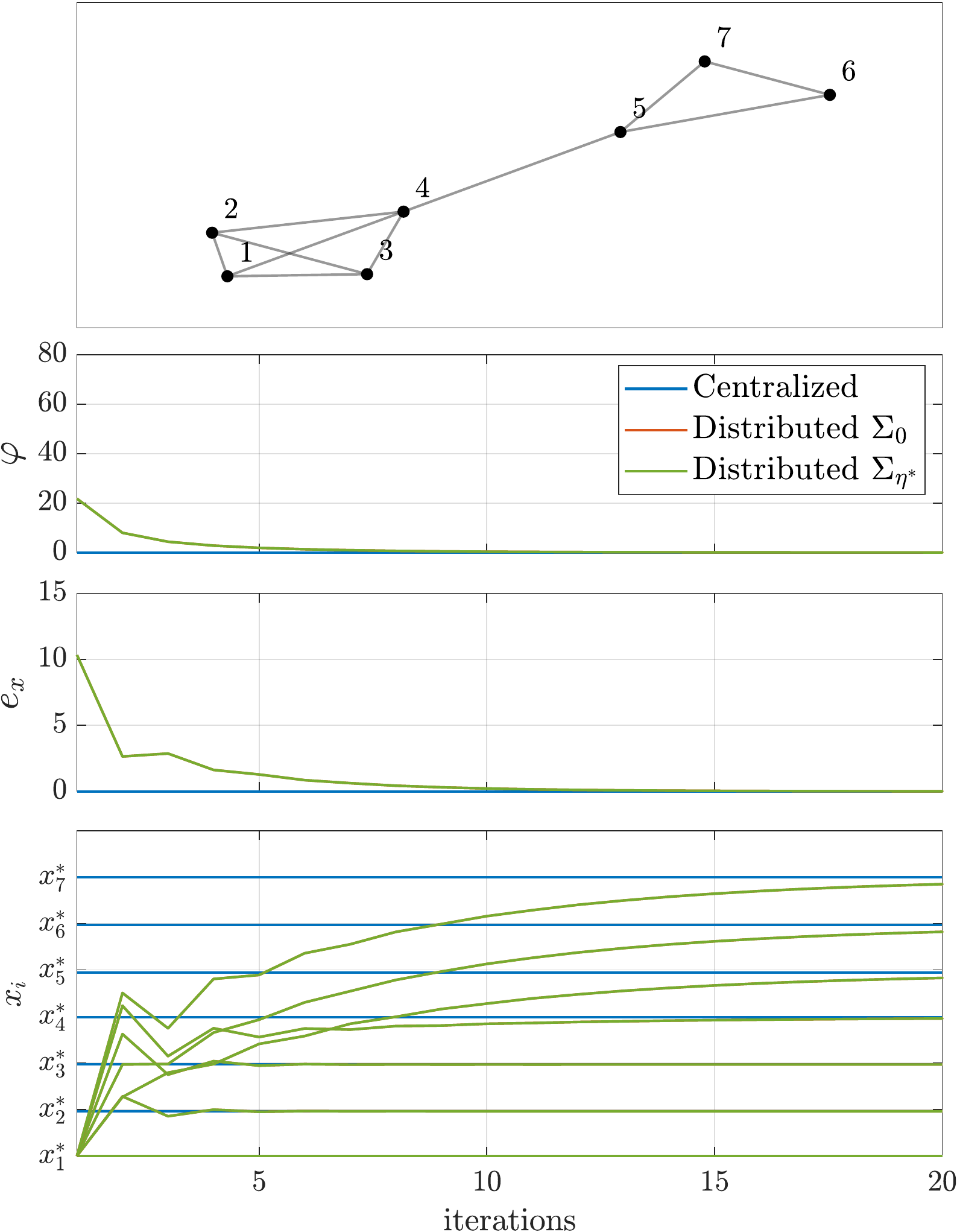}\label{fig:sim_3}}
		\caption{Simulation results considering three different network topologies composed by $n \in [5,8]$ agents: for each case, it is reported the related graph, the cost function, the performance index  and the  trend of the states (top to bottom boxes).} 
		\label{fig:sim_topology}
	\end{figure*} 

\subsection{Bipartite and Non-Bipartite Topologies}
	
Firstly, we consider the different graph-represented networks in the top panel of Fig.~\ref{fig:sim_topology}.  
In all the cases, the optimization~\eqref{eq:cost} is solved starting from zero initial conditions, i.e., ${x_i}(0)=0, \forall i \in \{1\ldots n\}$, using the centralized solution~\eqref{eq:centrSol} as the ground truth and stopping the iterative procedures after 20 iterations. 	The relative measurements are always assumed to be corrupted by additive uniform noise. 
The parameter $\eta$ is chosen as the optimized value, according to~\eqref{eq:etaast}.
	
To evaluate the convergence properties, a performance index is introduced, namely the \textit{mean squared error}. At each iteration $t$, this is defined as the squared distance
between the current estimates $\{ {x}_i (t)\}_{ i \in \{1 \ldots n\}}$ and the optimal values $\{ {x}_i^\ast\}_{ i \in \{1 \ldots n\}}$ given by the unique
centralized solution, i.e., 
	\begin{equation}
	\label{eq:psi}
	e_x(t) = \frac{1}{n}\sum_{i=1}^n ({x}_i(t)-x_i^\ast)^2.
	\end{equation}


The first test is conducted on an $n=8$ agent network represented by the regular non-circulant graph in the first plot in Fig.~\ref{fig:sim_1}. 
From the second and third plot in Fig.~\ref{fig:sim_1}, one can observe that both the cost functional and the performance index  converge to zero in less than five iterations independently on the distributed solution adopted. In addition, since the considered topology is not bipartite, also the state converges toward the equilibrium (bottom plot in Fig.~\ref{fig:sim_1}). 
Similar results are obtained when a non-regular small-world topology with $n=7$ nodes is evaluated (Fig.~\ref{fig:sim_3}). In this case, the behaviors of the schemes~\eqref{eq:distrUpdateVector1} and~\eqref{eq:distrUpdateVector_eta} are exactly the same due to the fact that $\varsigma_\mathbfcal{L} \leq 1$ and thus $\eta^\ast=0$.  
Differently, Fig.~\ref{fig:sim_2} illustrates the case of an $n=5$ agents network associated to a non-regular bipartite graph, showing different results in terms of convergence w.r.t. the other scenarios. Indeed, in this case, the adoption of model $\Sigma_0$ does not ensure the convergence and the tuning of parameter $\eta$ is necessary to reach a solution. 
	


\begin{table*}[t!]
	\centering
	\renewcommand\arraystretch{1.4}
	\begin{tabular}{c||c|c|c|c||c|c|c|c|}
		\cline{2-9}
		&
		{$K_{36}$} &
		{$C_{36}(1,2)$} &
		{$\mathcal{R}_{36}(3)$} &
		{$\Gamma_{36}(3)$} &
		{$G(36,0.1)$} &
		{$G(36,0.2)$} & 
		{$G(36,0.4)$} & 
		{$G(36,0.8)$} \\ \hline
		\multicolumn{1}{|c||}{$\phi$} & 
		1 & 9 & 7 & 6 & 
		6 & 6 & 6 & 5\\ \hline
		\multicolumn{1}{|c||}{$\eta^*$} & 
		0.0278 & 0 & 0 & 0.0557 & 
		0 & 0 & 0.0111 & 0.0182\\ \hline
		\multicolumn{1}{|c||}{$\mathfrak{r}_0$} & 
		0.0286 & 0.9623 & 0.9359 & 1 & 
		0.9262 & 0.8561 & 0.8317 & 0.8025\\ \hline
		\multicolumn{1}{|c||}{$\mathfrak{r}_{\eta^*}$} & 
		0 & - & - & 0.8885 & 
		- & - & 0.8114 & 0.7696\\ \hline
	\end{tabular}
	\caption{Convergence parameters of the considered networks (w.r.t. the topologies of Figs.~\ref{fig:regular_networks}-\ref{fig:ER_networks}(a)-(d)).}
	\label{tab:largescaleresults}
\end{table*}
%

\begin{figure}[t!]
	\centering
	\subfigure[Complete topology $K_{36}$]{\includegraphics[width=0.45\columnwidth]{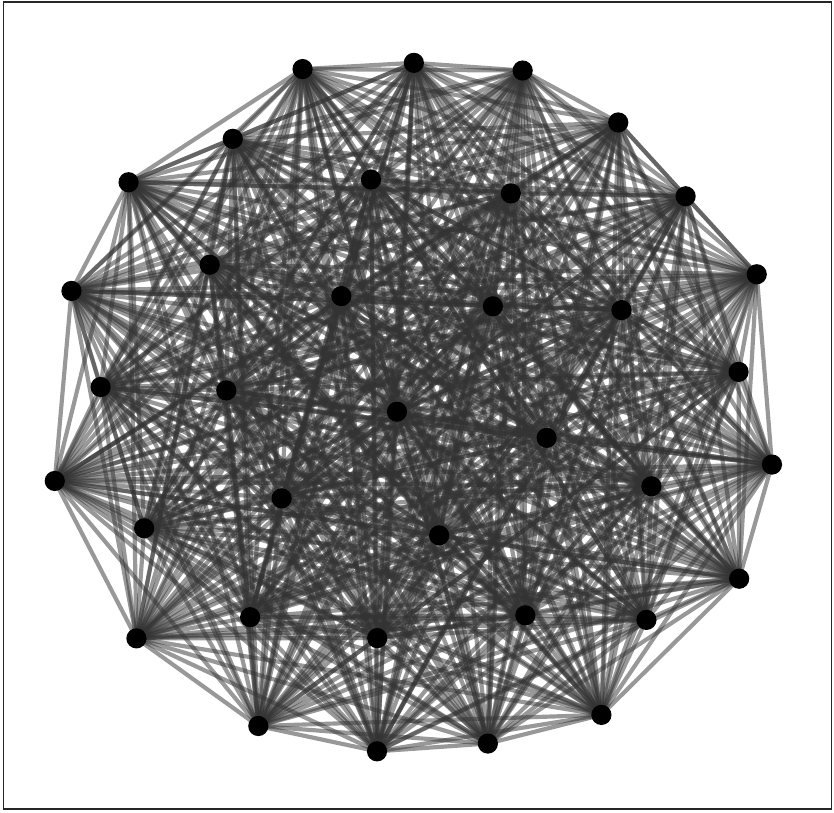}\label{fig:compl_graph}}\hspace{0.2cm}
	\subfigure[Circulant topology $C_{36}(1,2)$]{\includegraphics[width=0.45\columnwidth]{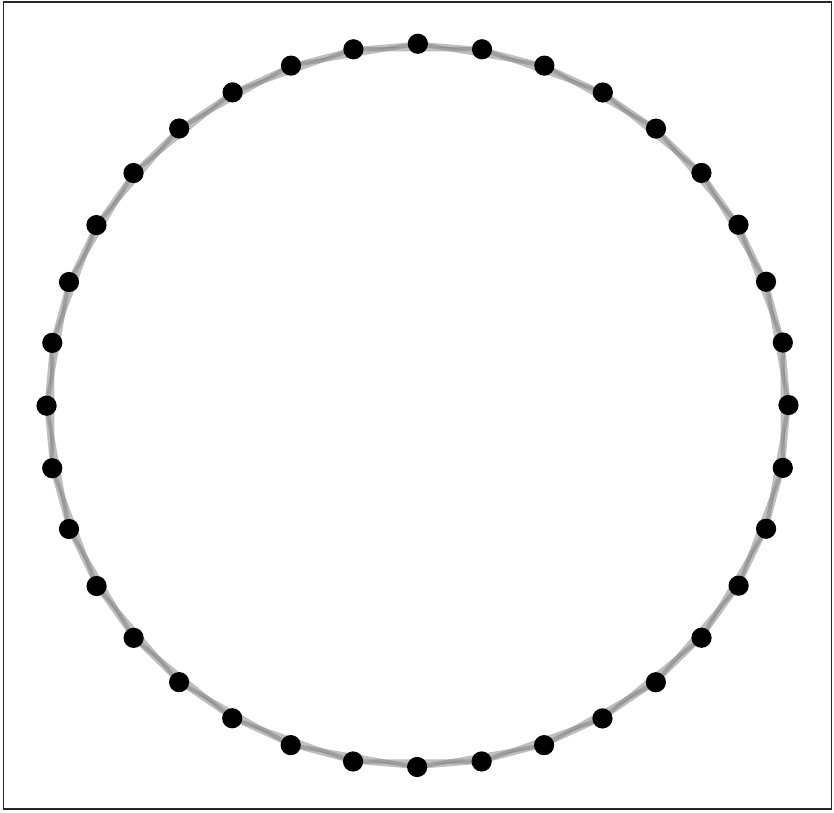}\label{fig:circ_graph}}\\
	\subfigure[Ramanujan topology $\mathcal{R}_{36}(3)$]{\includegraphics[width=0.45\columnwidth]{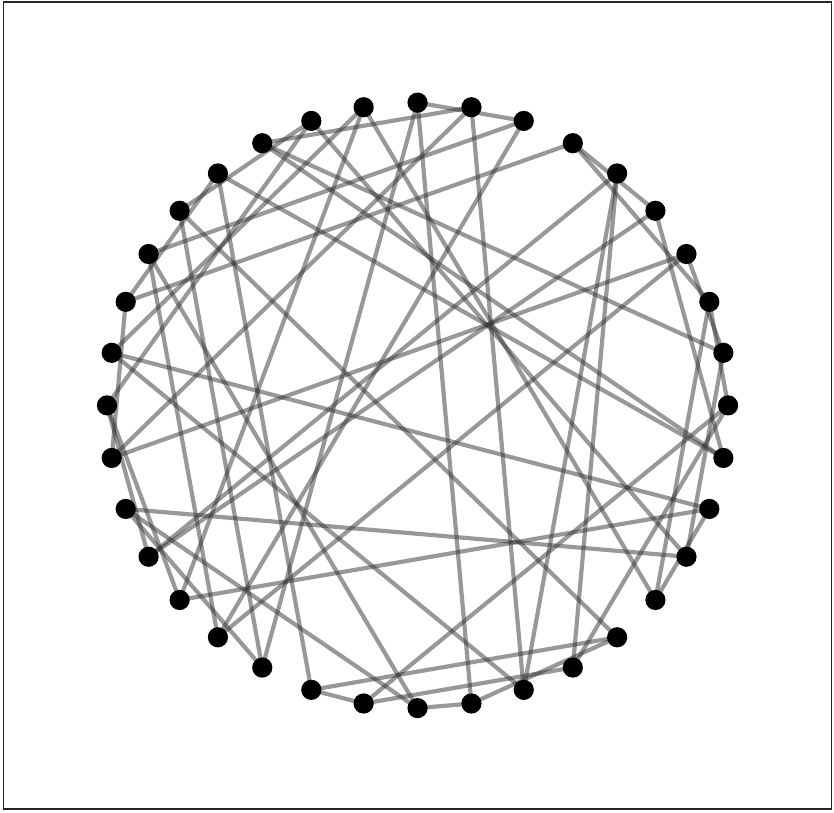}\label{fig:compl_graph}} \hspace{0.2cm}
	\subfigure[Cayley topology $\Gamma_{36}(3)$]{\includegraphics[width=0.45\columnwidth]{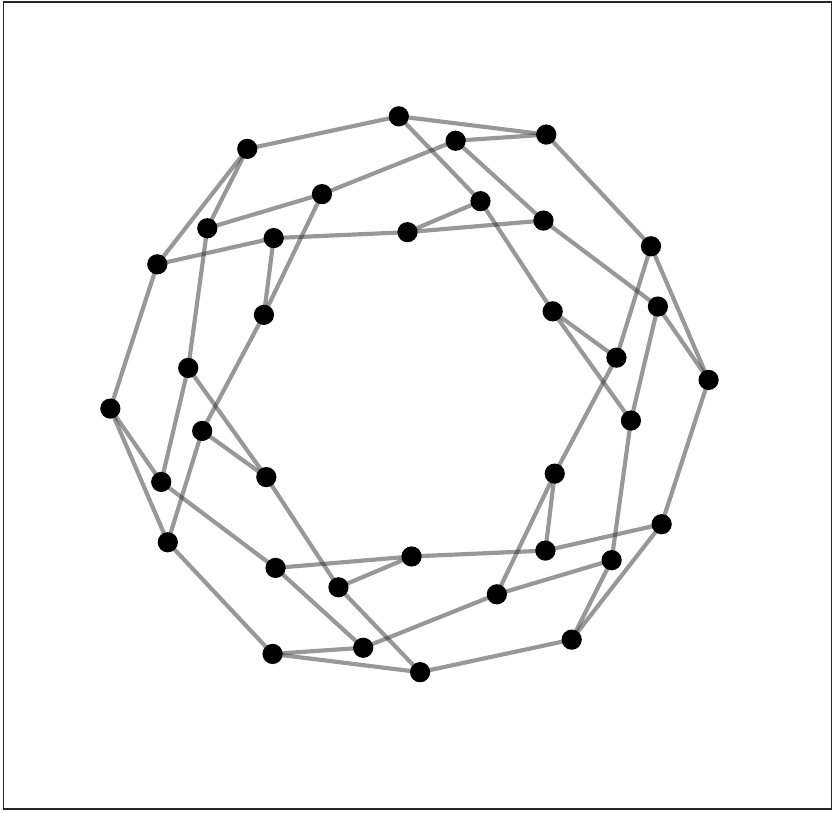}\label{fig:compl_graph}}
	\subfigure[Cost function trend]{\includegraphics[width=1\columnwidth, trim = {25 310 25 310}, clip]{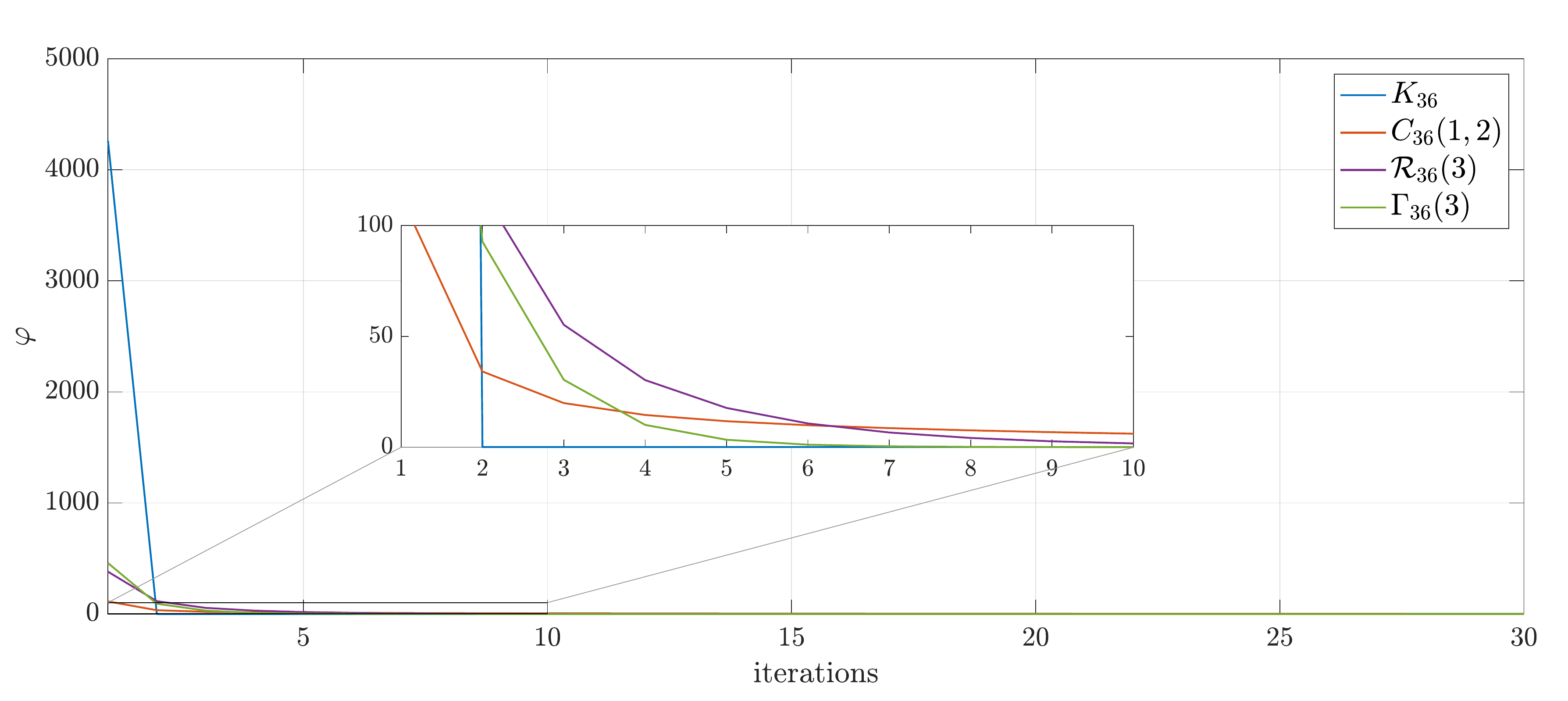}\label{fig:cost_reg}}
	\caption{Simulation results considering four different regular network topologies composed by $n=36$ agents: for each case, it is reported the related graph (top row) and the cost function by considering the distributed model $\Sigma_\eta$~\eqref{eq:distrUpdateVector_eta} (bottom plot).\\} 
	\label{fig:regular_networks}	
\end{figure}

\begin{figure}[t!]
	\centering
	\subfigure[$p=0.1$]{\includegraphics[width=0.45\columnwidth]{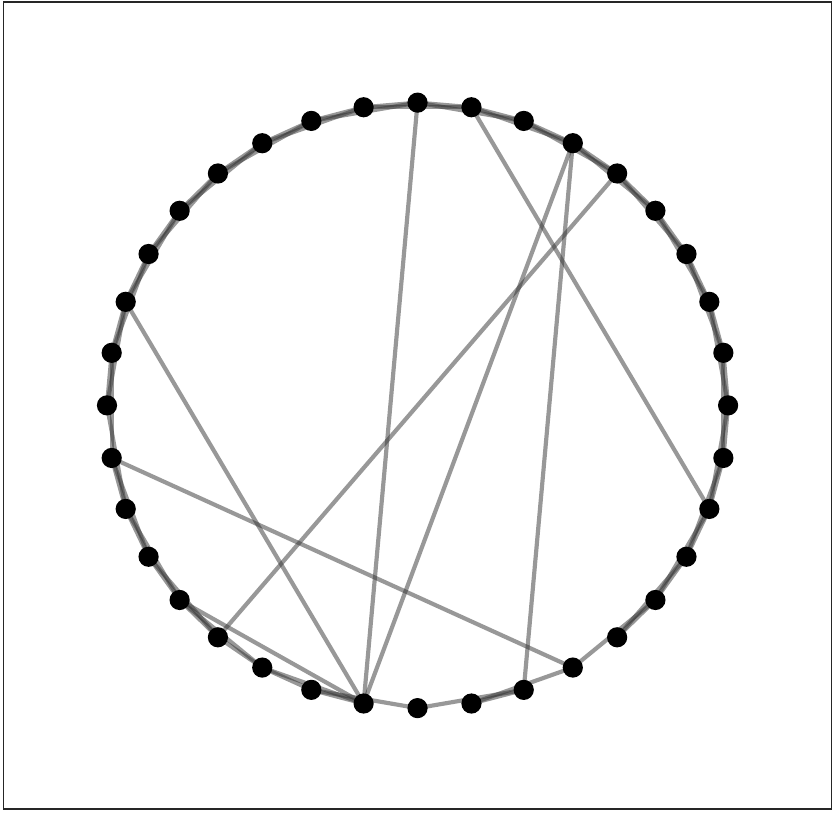}\label{fig:ER01_graph}}\hspace{0.2cm}
	\subfigure[$p=0.2$]{\includegraphics[width=0.45\columnwidth]{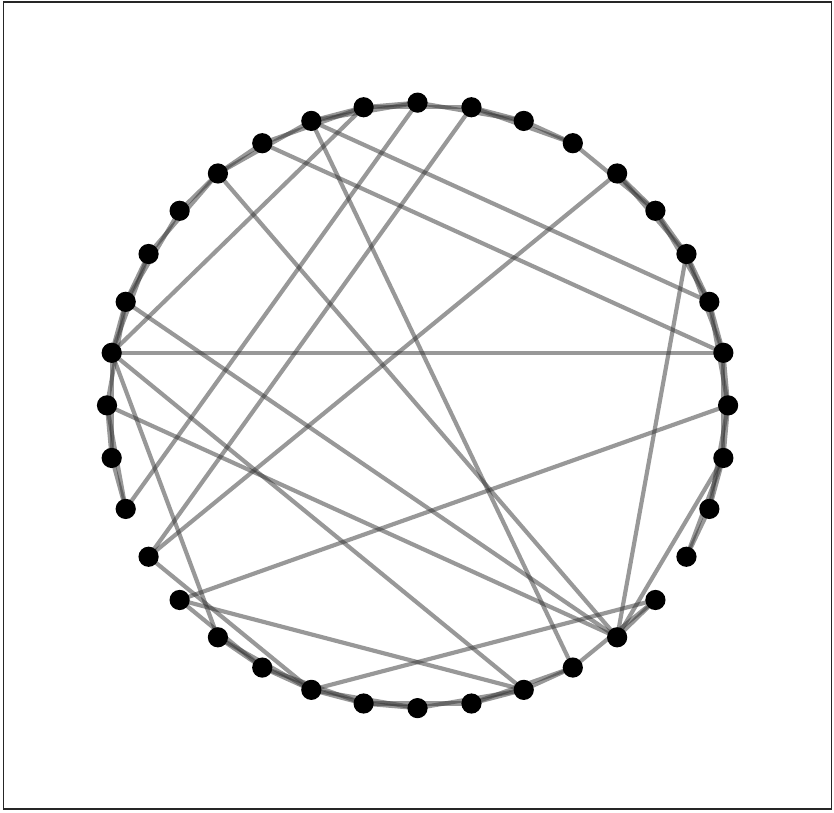}\label{fig:ER02_graph}}\\
	\subfigure[$p=0.4$]{\includegraphics[width=0.45\columnwidth]{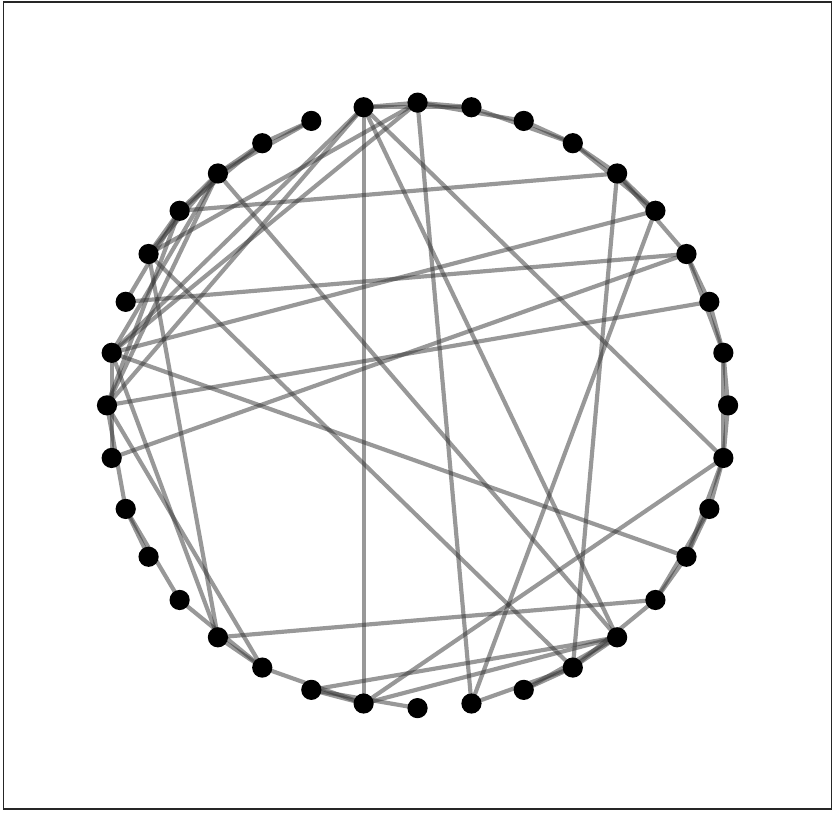}\label{fig:ER03_graph}} \hspace{0.2cm}
	\subfigure[$p=0.8$]{\includegraphics[width=0.45\columnwidth]{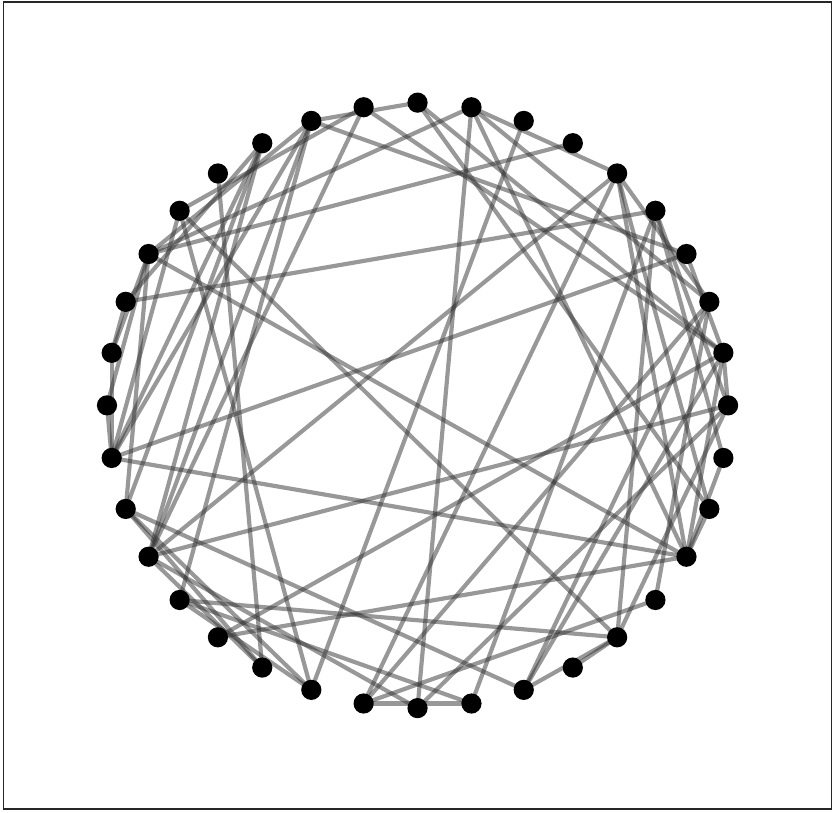}\label{fig:ER04_graph}}
	\subfigure[Cost function trend]{\includegraphics[width=1\columnwidth, trim = {25 310 25 310}, clip]{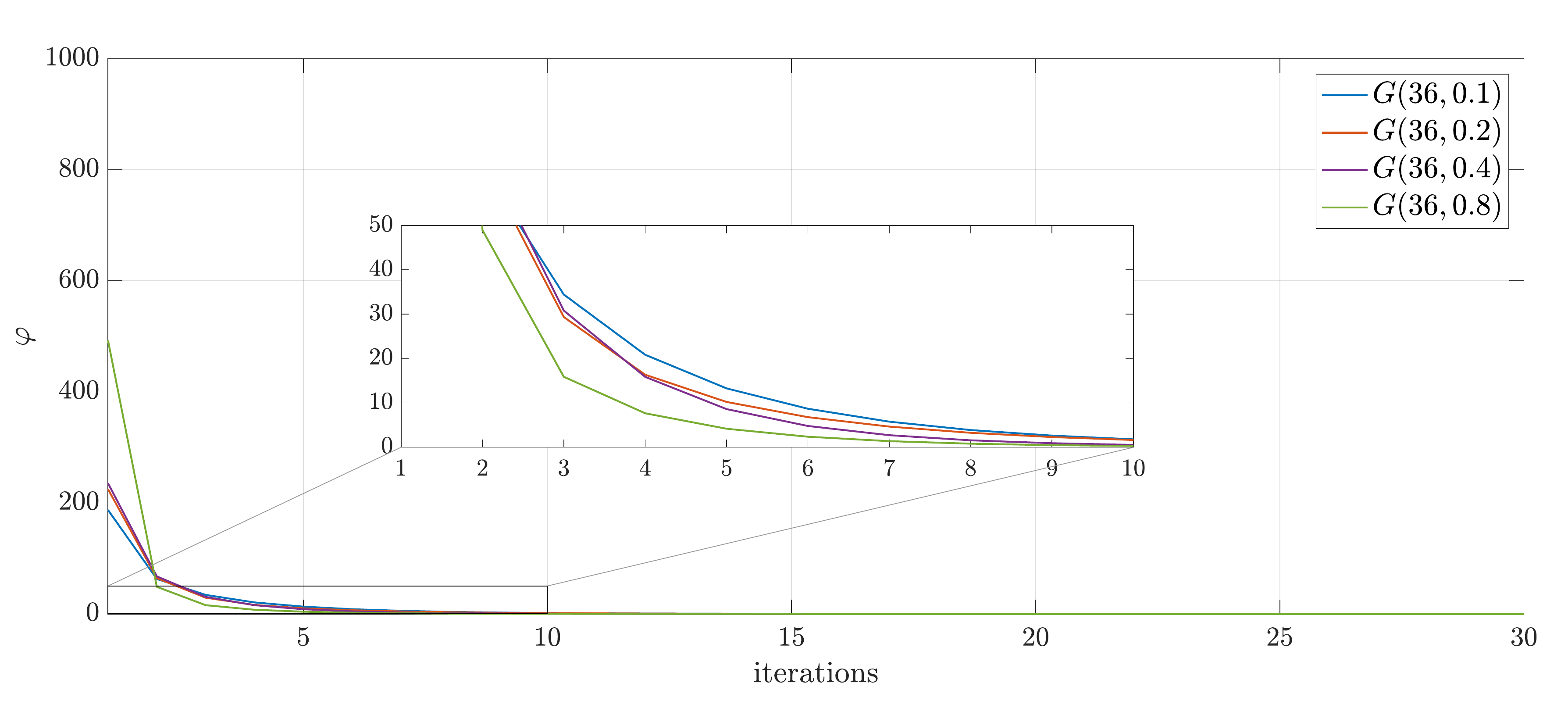}\label{fig:cost_ER}}
	\caption{Simulation results considering four different Erd\"os-R\'enyi network topologies composed by $n=36$ agents obtained by varying the edge connection probability: for each case, it is reported the related graph (top row) and the cost function by considering the distributed model $\Sigma_\eta$~\eqref{eq:distrUpdateVector_eta} (bottom plot).} 
	\label{fig:ER_networks}		
\end{figure}

\subsection{Large-scale Topologies}

We evaluate now some network topologies, all composed by $n=36$ agents. In detail, we consider the following regular networks (shown in Fig.~\ref{fig:regular_networks}(a)-(d)):
\begin{itemize}
\item complete graph $K_{36}$;
\item circulant graph $C_{36}(1,2)$, wherein each node is linked to the four closest neighbors;
\item Ramanujan graph ${\cal R}_{36}(3)$, with three neighbors per node;
\item Cayley graph ${\Gamma}_{36}(3)$, with three neighbors per node;
\end{itemize}  
and a set of Erd\"os-R\'enyi graphs $G(36, p)$, generated with different edge connection probability $p=\{0.1,0.2,0.4,0.8\}$
 (Fig.~\ref{fig:ER_networks}(a)-(d)).

For all the cases, the trend of the cost function $\varphi$ when the distributed model $\Sigma_\eta$ is considered is shown in the bottom panels of Figs.~\ref{fig:regular_networks}-\ref{fig:ER_networks}. In particular, the parameter $\eta$ is always set to its optimal value according to~\eqref{eq:etaast}. 
The simulation results confirm and highlight the theoretical findings and are summarized in Tab.~\ref{tab:largescaleresults}, where the following parameters are reported: graph diameter $\phi$, optimal value $\eta^\ast$, convergence rate $\mathfrak{r}_0$ for model $\Sigma_0$, and optimal convergence rate $\mathfrak{r}_{\eta^\ast}$ for model $\Sigma_\eta$. 
In this respect, some considerations are in order:

\subsubsection{regular networks}
For the complete topology $K_{36}$, we have that $\eta^\ast = n^{-1}= 0.0278$ as discussed in Sec.~\ref{sec:convergence_analysis_eta}: it is interesting to note that although fully connected the network needs memory (self-loops of $\Sigma_\eta$) in order to reach convergence in just one step (blue line in Fig.~\ref{fig:regular_networks}-(e)), in correspondence to $\mathfrak{r}_{\eta^\ast}=0$.\\
Instead, for both the networks represented by circulant graph $C_{36}(1,2)$ and $\mathcal{R}_{36}(3)$ it holds that $\eta^\ast=0$,~namely the model $\Sigma_0$ represents the best option to solve problem~\eqref{eq:cost}. 
Here, the convergence speed is slower for the circulant topology: $\varphi$ approaches zero in almost $30$ iterations (orange line) versus the $20$ needed in the Ramanujan graph.\\ 
An interesting case is that of the Cayley topology depicted in Fig.~\ref{fig:regular_networks}(d), since this network reveals to be bipartite ($\mathfrak{r}_0=1$): to attain convergence, model $\Sigma_\eta$ has to be employed. The simulation results confirm a continuous direct dependence between the $\eta$ parameter and the eigenvalue locations (hence the convergence rate), which actually allows to tune the convergence behavior and in the specific case to reach optimal performance in less than $10$ iterations.

\subsubsection{Erd\"os-R\'enyi networks}
In  all the four realizations of the Erd\"os-R\'enyi networks shown in Fig.~\ref{fig:ER_networks}, the convergence is always ensured, since non-bipartite graphs are obtained. In particular, for two networks it holds that $\eta^\ast=0$, namely through model $\Sigma_0$ the optimization problem~\eqref{eq:cost} is solved at best, while for the other two, the tuning of $\eta$ allows to reach better convergence performances w.r.t. the  model $\Sigma_0$.

%

\section{Conclusions}
	\label{sec:conclusions}
	
This paper focuses on the optimal (scalar) state estimation for the elements of a multi-agent system characterized by a set of noisy relative measurements. 
The problem is analytically solved in the convex minimization framework and two state-space models are derived for the update of the states. Particular attention is posed on the convergence properties of these update schemes, resting upon the spectral analysis of a class of stochastic matrices and related to the network topology. 
Numerical simulations support the theoretical considerations on the convergence performance, considering different scenarios, namely ring, bipartite and non-bipartite networks, and a variety of large-scale networks. 

In the future, the extension to multi-dimensional case is envisaged (this entails the determination of gradient-descent procedure over a proper manifold). Moreover, the performances of the spectral-based approach can be compared with other  optimized distributed estimation strategies.

%
%
%



%
	\bibliographystyle{IEEEtran}
	\bibliography{biblio}

\end{document}